# Regularised Iterative Generalised Least Squares with Optimal Selection of the Hyper-Parameter for Identifying Nonlinear Phenomenological Models

M. Cary[1] (Email: m.cary@lboro.ac.uk)

Charles Bokor[2] (Email: charlesbokor@outlook.com)

[1] *Department of Aeronautical and Automotive Engineering, Loughborough University, Loughborough, LE11 3TU, UK*

[2] *School of Engineering, Computing and Mathematics, Oxford Brookes University, Oxford OX33 1HX, UK*

## Abstract

In some fields currently dominated by empirical approaches, such as state of health *(SoH)* prediction for lithium-ion batteries, phenomenological models motivated by quasi-physical thinking contain parameters to be estimated from experimental data. Often the structure of such models yields fully or partially confounded parameters, which are difficult or even impossible to estimate reliably. To preserve the desired model formulation and simultaneously improve the numerical conditioning for the problem we introduce a ridge regression scheme. An automated method is provided, based on information theoretic measures of model performance, which optimises the ridge regression hyper-parameter at each iteration. The formulae presented require fixed point iteration to solve for the hyper-parameter. Given a suitable starting value, analysis demonstrates convergence is very rapid. The optimal hyper-parameter selection mechanism is incorporated within an efficient regularised iterative generalised least squares mechanism, capable of fitting both heteroscedastic and serially correlated data as required. Simulation confirms the efficacy of the overall method.

## 1 Preamble

The operation of Lithium-ion batteries creates ageing and degradation, which needs to be measured or predicted to keep the battery safe. Online state of health $(SoH)$ estimation is a key task of practical automotive battery management systems. Essentially $SoH$ is a metric to estimate the ageing level of batteries, which includes capacity fade and power degradation. Particularly, it is the key metric to determine when a battery should be replaced. Phenomenological models for $SoH$ estimation of Lithium-ion batteries rely heavily on quasi-physical motivations. Often these models are comprised of a power law like term multiplied by gains and Arrhenius-like exponential terms, which act as a form of rate constant [i, ii, iii]. Often the multiplicative nature of the model structure calls into question the apparent *identifiability* of the model.

A model is ***identifiable*** if it is theoretically possible to learn the true values of this model's underlying parameters after obtaining an infinite number of observations from it. To make things concrete let $P = \{P_\theta : \theta \in \Theta\}$ denote a statistical model, where the parameter space, $\Theta$, is finite-dimensional. The model $P$ is identifiable if the mapping $\theta \to P_\theta$ is one-to-one; *i.e.* $P_{\theta_1} = P_{\theta_2} \Rightarrow \theta_1 = \theta_2, \forall \theta_1, \theta_2 \in \Theta$. Identifiability of the model, in the sense of the invertibility of the mapping $\theta \to P_\theta$, is equivalent to being able to discern the true parameter vector if the model can be observed indefinitely.

Let $P$ be the general nonlinear regression model:

$$y = f(x, \theta) + e \quad, e = \boldsymbol{\mathcal{N}}\left(0, \sigma^2 v^2(f(x, \theta), \delta)\right) \tag{1}$$

Note although the error term is assumed Gaussian in nature, it is permitted to take on quite general structures incorporating phenomena such as heteroscedasticity and or serial correlation through the specification of the variance-covariance function $v^2(f(x, \theta), \delta)$. Expanding $f(x, \theta)$ as a first order Taylor series about the current estimates $\theta_k \in \mathbb{R}^p$ and defining $g = y - f(x, \theta_k)$, Equation (1) becomes:

$$g \approx J(x, \theta_k) \Delta\theta + e \tag{2}$$

Where $J(x, \theta_k) \in \mathbb{R}^{N \times p}$ is the model Jacobian matrix defined as $J(x, \theta_k) = \partial f(x, \theta_k)/\partial\theta$. Identifiability is closely linked with sensitivity analysis [iv, v, vi], and so to the properties of $J(x, \theta)$. Given that the experimental design is adequate, if the columns of the Jacobian matrix are linearly independent then the parameter estimates can be uniquely estimated. This is equivalent to stating that the matrix $\left(J(x, \theta)^T J(x, \theta)\right)$ is invertible.

Vajda *et al* [vii] investigated the *quantitative identifiability* of model parameters for a flow reactor employed to pyrolyze methane at three constant temperatures. Their approach employed principal component analysis [viii] to determine the number of ''large'' eigenvalues of the Fisher Information Matrix. The elements of the corresponding eigenvectors were used to suggest combinations of parameters that could be used to re-parameterize and simplify the model; the number of large eigenvalues corresponding to the number of estimable parameters in the simplified model. Simplifying the model seems at odds with the underlying physical thinking often used to formulate a Lithium-ion battery $SoH$ model, where certain parameters may account for known ageing mechanisms. Indeed, obtaining estimates of the associated parameters to rank the relative significance of individual mechanisms may be the primary goal of the study.

Vajda's approach was limited to small scale models, with relatively few parameters. Yao *et al* [iv] define an approach capable of dealing with larger scale problems. Their example contains up to 50 parameters. The algorithm developed is analogous to calculating the alias matrix in linear regression problems [ix] and relies on the fact that the residual vector is orthogonal to the expectation plain. Each column added to the simplified scaled Jacobian is thus orthogonal to any existing columns. Again, the main concern is that the resulting model is simplified, and the original physical formulation altered.

In situations where the columns of $J(x, \theta)$ are nearly colinear, the conditioning of the Jacobian is very poor, resulting in very imprecise parameter estimates. In addition, the parameter estimates are very sensitive to the peculiarities of the training data, implying that a small perturbation in the data can lead to a large change in the parameter estimates [x]. One way of converting a problem from ill-posed to well-posed is to introduce a regularisation method. The basic idea is to confine the problem to a restricted set of parameters using a regularisation functional $\mathcal{R}(\theta)$. Typically, this is achieved by introducing a constraint of the form $\mathcal{R}(\theta) \leq c$ to the usual log likelihood function, $L(\theta)$, such that:

$$arg \min_{\theta} \left( L(\theta) + \frac{\lambda}{2} \mathcal{R}(\theta) \right) \tag{3}$$

This is reminiscent of the method of Langrage multipliers. The regularisation parameter $(\lambda)$, or hyper-parameter, implicitly defines a structure on the possible models by constraining the model. Roughly speaking, low values of $\lambda$ in Equation (3) correspond to high values of $c$; *i.e.* a weak constraint. Alternatively, a large value for the hyper-parameter, corresponding to a low value for $c$, implies greater importance to $\mathcal{R}(\theta)$ in the minimisation of the cost function. Thus, a large value for $\lambda$ represents a more severe constraint. Hence, we see the regularised cost minimisation problem represents a trade-off between model fidelity, minimising $L(\theta)$, and constraining the model to be a member of a small compact subset, $\mathcal{R}(\theta)$. Furthermore, the balance between fitting the data and satisfying the constraint is controlled by the regularisation parameter.

In practice, minimisation of the regularised cost function is carried out iteratively using a gradient based optimisation method. In this paper we are concerned with optimal selection of the hyper-parameter at each iteration of this process. In this way, the fidelity versus constraint level trade-off can be handled automatically as the iterative parameter search proceeds. This is accomplished by developing a fixed-point iteration scheme to be carried out at each iteration of the optimisation scheme. These are founded on novel re-estimation formulae presented in this paper.

The paper is organised as follows. In section 2 we introduce the information theoretic model selection measures upon which the optimal hyper-parameter estimation process is founded. Both information theoretic measures are functions of the relevant loglikelihood, and in section 3 we derive the appropriate profiled likelihood necessary for the derivation of the re-estimation formulae for the hyper-parameter. For continuity, section 4 we provide expressions for the derivatives of the profile loglikelihood and model degrees of freedom with respect to the hyper-parameter. These are necessary for the derivation of the re-estimation formulae presented in section 5 and derived in detail in Appendix B. Convergence criterion for the iterative algorithm are also given here. Section 6 details the regularised iterative generalised least squares algorithm utilised for the identification of both the fit and covariance model parameters. An example analysis of simulated data follows in section 7. The application considered is the Martinez-Laserna model for battery state of health estimation. The multiplicative structure of the model yields particularly poor identifiability and regularisation proves essential to obtain reasonable parameter estimates. In section 8, we comment on the convergence rate of the iterative process and its dependency on the initial derivative of the re-estimation function with respect to the hyper-parameter. Finally, in section 9 we state conclusions relating to the efficacy of the procedure.

## 2 Model Selection Measures

Various measures appear in the literature, which permit comparison among competing statistical models [xi, xii, xiii, xiv]. Orr [xv, xvi] provides a method for optimally estimating the hyper-parameter in regularised radial basis function network applications. His approach is based on the well-known generalised cross-validation ($GCV$) criterion [xiv]. The main justification for $GCV$ as a model selection criterion is conveniently summarised in their $GCV$ theorem [xiv, xvii, xviii]. As discussed by Eubank [xvii] the GCV theorem can be interpreted as stating that, in certain circumstances, GCV is nearly an unbiased estimator of the prediction risk.

Essentially, Orr presents a *re-estimation* method for optimally determining $\lambda$ as a new basis function is added to the model. By differentiating the relevant $GCV$ criterion with respect to the hyper-parameter, setting the result to zero and solving for $\lambda$ he derives an expression for updating the regularisation parameter as each new basis function is added to the network. As $\lambda$ appears on both sides of this expression, iteration is required to determine the updated value.

There are several issues associated with Orr's approach. Firstly, it relates to ordinary least squares estimation only. Weighted fitting approaches are not accounted for. Secondly, the measure has units which means that models created in transformed scales cannot be compared. Finally, Orr's parameter count for the model neglects the covariance parameter; $\sigma^2$ in the least squares case.

In contrast, we base our approach on two popular information theoretic criteria: Sugiura's small sample Akaike Information Criterion ($AIC_c$) [xiii] and also Schwarz's Bayesian Information Criterion [xix] ($BIC$). These are defined as:

$$AIC_c = 2L(\theta) + \frac{2\gamma(\gamma+1)}{N-\gamma-1} \tag{4}$$

$$BIC = 2L(\theta) + \gamma log(N) \tag{5}$$

Where $L(\theta)$ is the negative loglikelihood, $N$ the number of data points and $\gamma$ the total number of estimable parameters in the model specification. We derive formulae based on both measures and compare their respective performance in later sections. Notice the full version of both $BIC$ and $AIC_c$, as given in Equations (4) and (5), are based on the negative loglikelihood, which is unitless. Therefore, these measures permit comparisons among models computed in both transformed and natural unit scales. The situation is analogous to the well know Box-Cox method [xx] for determining an optimal transformation in linear regression problems. In addition, by employing the full likelihood we can account for heteroscedastic and or serially correlated data.

For either measure, the estimable parameter count necessarily includes both fit and covariance model parameters. In addition, the term *estimable* deserves some clarification, particularly for regularised models. Following Burnham and Anderson [xxi], by this we imply parameters that are uniquely identifiable from the data. One reason a parameter can be non-estimable is due to inherent confounding in the model. Under these circumstances, two or more parameters cannot be uniquely identified from the data. This is obvious in multiplicative structures involving products of coefficients. Such formulations are relatively common in phenomenological models for lithium-ion battery $SoH$ estimation.

## 3 Approximations to the Log-Likelihood and the Ridge Estimator

We begin by assuming Gaussian distributed errors, at the $k^{th}$ iteration, according to $\boldsymbol{e}_k = \boldsymbol{\mathcal{N}}(\mathbf{0}, \sigma^2 W)$. Where the general covariance matrix, $\sigma^2 W$, may incorporate heteroscedastic and or serially correlated errors. Further, let $\theta \in \mathbb{R}^p$ denote the vector of parameters to be identified and $\boldsymbol{f}(\theta)$ the model to be fitted. Under these circumstances, the corresponding negative loglikelihood is:

$$L = \frac{N}{2} log(2\pi) + \frac{N}{2} log(\sigma^2) + \frac{1}{2} log(\|W\|) + \frac{1}{2\sigma^2}\big(\boldsymbol{y} - \boldsymbol{f}(\theta)\big)^T W^{-1}\big(\boldsymbol{y} - \boldsymbol{f}(\theta)\big) \quad (6)$$

In general, there is no closed form solution for the estimation of $\theta$ and we employ iterative methods. For the $k^{th}$ iteration, we expand $\boldsymbol{f}(\theta)$ about $\theta_k$ using a first-order Taylor series. Thus:

$$\boldsymbol{f}(\theta) \sim \boldsymbol{f}(\theta_k) + J(\theta_k)\Delta\theta \quad (7)$$

Where $J(\theta_k) = \partial \boldsymbol{f}(\theta_k)/\partial\, \theta_k \in \mathbb{R}^{N \times p}$. Substituting (7) into (6), dropping the notational dependency on $\theta_k$ and writing $\boldsymbol{g} = \boldsymbol{y} - \boldsymbol{f}(\theta)$:

$$L \sim \frac{N}{2} log(2\pi) + \frac{N}{2} log(\sigma^2) + \frac{1}{2} log(\|W\|) + \frac{1}{2\sigma^2}(\boldsymbol{g} - J\Delta\theta)^T W^{-1}(\boldsymbol{g} - J\Delta\theta) \quad (8)$$

As $W$ is real and positive definite, it may be factored as $W = C^T C$, where $C$ is upper triangular. Using standard results on matrix inverses [xxii], it can be shown that $W^{-1} = C^{-1}C^{-T}$. Further, as $C$ is triangular, $log(\|W\|) = 2\sum_{i=1}^{N} log(c_{ii})$, where $c_{ii}$ is the $i^{th}$ term on the leading diagonal of $C$. Applying these results and writing $\boldsymbol{q} = C^{-T}\boldsymbol{g}$ and $Z = C^{-T}J$, yields:

$$L \sim \frac{N}{2} log(2\pi) + \frac{N}{2} log(\sigma^2) + \sum_{i=1}^{N} log(c_{ii}) + \frac{1}{2\sigma^2}(\boldsymbol{q} - Z\Delta\theta)^T(\boldsymbol{q} - Z\Delta\theta) \quad (9)$$

We now introduce a ridge regression term into (10) to form:

$$R \sim \frac{N}{2} log(2\pi) + \frac{N}{2} log(\sigma^2) + \sum_{i=1}^{N} log(c_{ii}) + \frac{1}{2\sigma^2}(\boldsymbol{q} - Z\Delta\theta)^T(\boldsymbol{q} - Z\Delta\theta) + \frac{\lambda}{2\sigma^2}\Delta\theta^T\Delta\theta \quad (10)$$

Differentiating (11) w.r.t. $\Delta\theta$, setting the result to zero and solving for $\Delta\theta$ yields after some algebra:

$$\Delta\theta = (Z^T Z + \lambda I)^{-1} Z^T \boldsymbol{q} = A^{-1} Z^T \boldsymbol{q} \quad (11)$$

Substituting (12) into (10) and factorising:

$$L \sim \frac{N}{2} log(2\pi) + \frac{N}{2} log(\sigma^2) + \sum_{i=1}^{N} log(c_{ii}) + \frac{1}{2\sigma^2} \boldsymbol{q}^T \big(I - S(\lambda)\big)^2 \boldsymbol{q} \quad (12)$$

Where, $S = ZA^{-1}Z^T$ is the so-called smoother matrix. Note $S$ is symmetric, so $S = S^T$. Additionally, the effective degrees of freedom associated with the model, $\gamma$, is given by $\gamma = tr(S)$ [xv, xxiii]. We use (13) as the basis for optimal re-estimation of the hyper-parameter in section 5. In addition, we can exploit the fact that the likelihood is conditionally linear in the variance scale parameter $\sigma^2$. Consequently, a closed form solution for the estimate of $\sigma^2$ exists, which can be shown to be:

$$\sigma^2 = \frac{\boldsymbol{q}^T (I - S)^2 \boldsymbol{q}}{N} \quad (13)$$

Substituting Equations (14) into (13) yields the corresponding *profile likelihood* $(L_p)$:

$$L_p \sim \frac{N}{2} log(2\pi) + \frac{N}{2} log\left(\frac{\boldsymbol{q}^T \big(I - S(\lambda)\big)^2 \boldsymbol{q}}{N}\right) + \sum_{i=1}^{N} log(c_{ii}) + \frac{N}{2} \quad (14)$$

Equation (14) is used as the basis for the covariance model parameters as detailed in section 6 and for optimally selecting the hyper-parameter as described in section 5.

# 4 Derivatives of the Smoother Matrix, $\gamma$ and the Profile Loglikelihood w.r.t. $\lambda$

Both information theoretic measures involve the derivative of $L_p$ and $\gamma$ with respect to $\lambda$. This necessarily involves formulae for the corresponding derivatives of $S$, $tr(S)$ and $S^2$. These are derived in appendix A, and only the results stated here for continuity. Hence:

$$\frac{\partial S}{\partial \lambda} = -ZA^{-2}Z^T \quad (15)$$

$$\frac{\partial S^2}{\partial \lambda} = 2S\frac{\partial S}{\partial \lambda} = -2ZA^{-2}Z^T + 2\lambda ZA^{-3}Z^T \quad (16)$$

Similarly, recalling $\gamma = tr(S)$:

$$\frac{\partial \gamma}{\partial \lambda} = tr(\lambda A(\lambda)^{-2} - A(\lambda)^{-1}) \quad (17)$$

Finally, using these results, it can be shown after some matrix algebra that the derivative of the loglikelihood, Equation (13), is given by:

$$\frac{\partial L_p}{\partial \lambda} = \lambda\left(\frac{N}{\sigma(\lambda)^2}\right)\boldsymbol{q}^T Z A(\lambda)^{-3} Z^T \boldsymbol{q} \tag{18}$$

## 5 Optimal Selection of the Hyper-Parameter Based on $BIC$ and $AIC_c$

In appendix B we derive the following fixed-point iteration formulae based on BIC and $AIC_c$:

$$\lambda_{BIC} = \frac{\sigma^2 log(N) tr\left(A^{-1} - \lambda A^{-2}\right)}{2N\boldsymbol{q}^T Z A^{-3} Z^T \boldsymbol{q}} \tag{19}$$

$$\lambda_{AIC_c} = \frac{\sigma^2\left[\frac{(2\gamma+1)}{(N-\gamma-1)} - \frac{\gamma(\gamma+1)}{(N-\gamma-1)^2}\right] tr\left(A^{-1} - \lambda A^{-2}\right)}{N\boldsymbol{q}^T Z A^{-3} Z^T \boldsymbol{q}} \tag{20}$$

Notice in Equations (19) and (20), for notational compactness, we drop the dependency of $\sigma$, $\gamma$ and $A$ on $\lambda$. Regardless, $\lambda$ appears on both sides of either expression. Consequently, we must resort to iterative methods to estimate the optimal $\lambda$. Furthermore, we use $h(\lambda)$ to denote the righthand side of (19) or (20). These represent a *fixed-point iteration scheme* [xxiv], with both equations in the form $h(\lambda) = \lambda$. A *fixed-point* is defined by:

***Definition***: If $h(\lambda)$ is defined on $[a, b]$ and $h(\lambda) = \lambda$ for some $\lambda \in [a, b]$, then the function $h(\lambda)$ is said to have the fixed-point $\lambda$ in $[a, b]$.

The following theorem guarantees convergence of the algorithm:

***Theorem***: Let $h(\lambda)$ be continuous in $[a, b]$, and suppose that $h(\lambda) \in [a, b]$ for all $\lambda \in [a, b]$. Further Let $h'(\lambda)$ exist on $(a, b)$ with $|h'(\lambda)| \leq k < 1$, for all $\lambda \in (a, b)$. If $\lambda_o \in [a, b]$, then the sequence defined by $\lambda_r = h(\lambda_{r-1})$, $r \geq 1$, will converge to the unique fixed-point $\lambda$ in $[a, b]$.

The necessary proof can be found in the reference cited. Further it follows that if the theorem is satisfied then a bound for the error involved in using $\lambda_r$ to approximate $\lambda$ is given by:

$$|\lambda_r - \lambda| \leq k^r max\{\lambda_o - a, b - \lambda_o\}, \forall r \geq 1. \tag{21}$$

The theorem requires $h'(\lambda)$, which we calculate numerically using the 5-point stencil approximation provided in [xxv]. This is accurate to $O(\Delta\lambda^4)$. Thus, the fixed-point iteration algorithm is:

1. Generate 6 logarithmically spaced values in the interval $\lambda \in [10^{-5}10^{0}]$ and calculate the corresponding $|dh(\lambda_0)/d\lambda|$. Select the $\lambda$ for which the following conditions are met:
    a. $min\{|dh(\lambda_0)/d\lambda|\} < 1$.
    b. $h(\lambda_0) \in [10^{-5}10^{0}]$.
    c. Denote this value by $\lambda_r$. Set $r = 0$.
2. Increment $r = r + 1$.
3. Calculate $\lambda_{r+1}$ using either Equation (19) or (20) as desired.
4. If $(|\lambda_{r+1} - \lambda_r|/\lambda_r \leq 0.0001)\ or(\ i \geq i_{max}) \rightarrow stop$. Else treating $\lambda_{r+1}$ as $\lambda_r$ return to step 2.

Having provided the re-estimation formulae and associated algorithm for optimal hyper-parameter selection, we now detail the full nonlinear regularised iterative generalised least squares ($RIGLS$) algorithm, with optimal $\lambda$ selection at each iteration.

# 6 A $RIGLS$ Algorithm with Optimal Hyper-Parameter Selection

Davidian and Giltinan [xxvi] provide a rationale for preferring iterative generalised least squares over joint maximum likelihood. Their argument is based on the improved robustness to both model misspecification and non-normality. Here, we extend their basic algorithm to the regularised case and for the profiled loglikelihood. The regularised iterative generalised least squares algorithm employed is as follows:

1. Obtain the initial estimates, $\theta_r$, by minimising the following nonlinear functional with respect to $\theta$. Re-estimate $\lambda$ at each iteration, by iterating equation (19) or (20) to convergence as required.
    - $arg \min_{\theta} R = \frac{1}{2}\left(y - f(\theta)\right)^T\left(y - f(\theta)\right) + \frac{\lambda}{2}\theta^T\theta$
2. Using $\theta_r$, form the residuals, $d_r = y - f(\theta_r)$. Using these, minimise the following functional with respect to $\delta$. Denote these estimates as $\delta_{r+1}$.
    - $arg \min_{\delta}\left(L_p = \frac{N}{2}log\left(\frac{d_r^T W^{-1}(\theta,\delta) d_r}{N}\right) + \frac{1}{2}log\|W(\theta,\delta)\| + \frac{N}{2}\ \right)$
3. Using $\delta_{r+1}$ construct the matrix $W^{-1}(\theta_r,\ \delta_{r+1}\ )$. Update estimates for $\theta$ by minimising the nonlinear function:
    - $arg \min_{\theta} R = \{\frac{1}{2}\left(y - f(\theta)\right)^T W^{-1}(\theta_r,\ \delta_{r+1}\ )\left(y - f(\theta)\right) + \frac{\lambda}{2}\theta^T\theta\}$
    - Re-estimate $\lambda$ at each iteration, by iterating equation (20) or (21) to convergence as required.
    - Denote the values at convergence by $\theta_{r+1}$.
4. Compute $\beth = 1 - \frac{\|\theta_{r+1}\|_{r+1}^2}{\|\theta_r\|_r^2}$. If $\beth > 0.0001$, then treating $\theta_{r+1}$ as $\theta_r$ return to step 2, else stop.

- Compute $\sigma^2 = \frac{d_r^T W^{-1}(\theta,\delta) d_r}{N}$.

The algorithm has been implemented in MATLAB and utilises the ***fmincon*** function from the MATLAB *optimisation toolbox*. To work as intended, it is necessary to supply analytical gradients to the ***fmincon*** routine, so that the hyper-parameter is calculated only once per iteration, as opposed to $p$-times if finite differences are utilised. Using standard results on matrix derivatives [xxii], it is shown in Appendix C that:

$$\frac{\partial R}{\partial \theta} = J^T(\theta) W^{-1}(f(x,\theta) - y) + \lambda\theta \tag{22}$$

Note, as required at step 1, $W = I \in \mathbb{R}^{N\times N}$. The authors have constructed a MATLAB package to implement the algorithm, which makes use of the *strategy* and *template* object oriented behavioural patterns defined by Gamma *et al* [xxvii].

# 7 Example Analysis

For example, consider the model proposed by Martinez-Laserna *et al* [i], which attempts to predict capacity fade, $Q_{loss}$:

$$Q_{loss} = \omega. exp\left(\frac{\beta_1}{T} + \beta_2 \overline{SOC}\right) Ah^z \tag{23}$$

With $Ah$ the accumulated charge throughput, $T$ the cell absolute thermodynamic temperature and $\overline{SOC}$ the average state of charge. The parameter vector, $\theta = (\omega, \beta_1, \beta_2, z)$, is to be estimated from the data. The multiplicative structure is common in such models for this application and results in extremely poor identifiability. The Martinez-Laserna model has been fitted to simulated data representing a cyclic ageing profile for a single cell. The advantage of utilising simulated data is estimates can be compared to the known true values, thus demonstrating the efficacy of the algorithms employed. The *true* model is:

$$Q_{loss} = 100 \times 0.4. exp\left(\frac{0.09}{(T+273.15)} + 0.02. SOC\right). Ah^{0.5} \tag{24}$$

In addition, the simulated data is contaminated with Gaussian heteroscedastic noise of the form $\mathcal{N}(0, \sigma^2. Q_{loss}^{2\delta})$ with $\sigma = 0.05$ and $\delta = 1$. Again, these quantities are to be estimated from the data. Fifty-one equally spaced points were generated for the interval $Ah \in [0,50]$. During fitting, both $Ah$ and $Q_{loss}$ are mapped onto the interval $[0, 1]$.

Figure 1 presents the usual diagnostic plots for a regularised ordinary least squares analysis. The hyper-parameter is optimised using the algorithm detailed in section 5 and is based on the $AIC_c$ performance measure. This is equivalent to performing step 1 only of the $RIGLS$ algorithm of section

6. We refer to this as the regularised ordinary least squares ($ROLS$) model. The heterogenous nature of the variance is apparent from the residual versus predicted plot, which exhibits a distinct increase in the variance scale with increasing predicted $Q_{loss}$. The optimal value of the hyper-parameter at convergence is 0.0074.

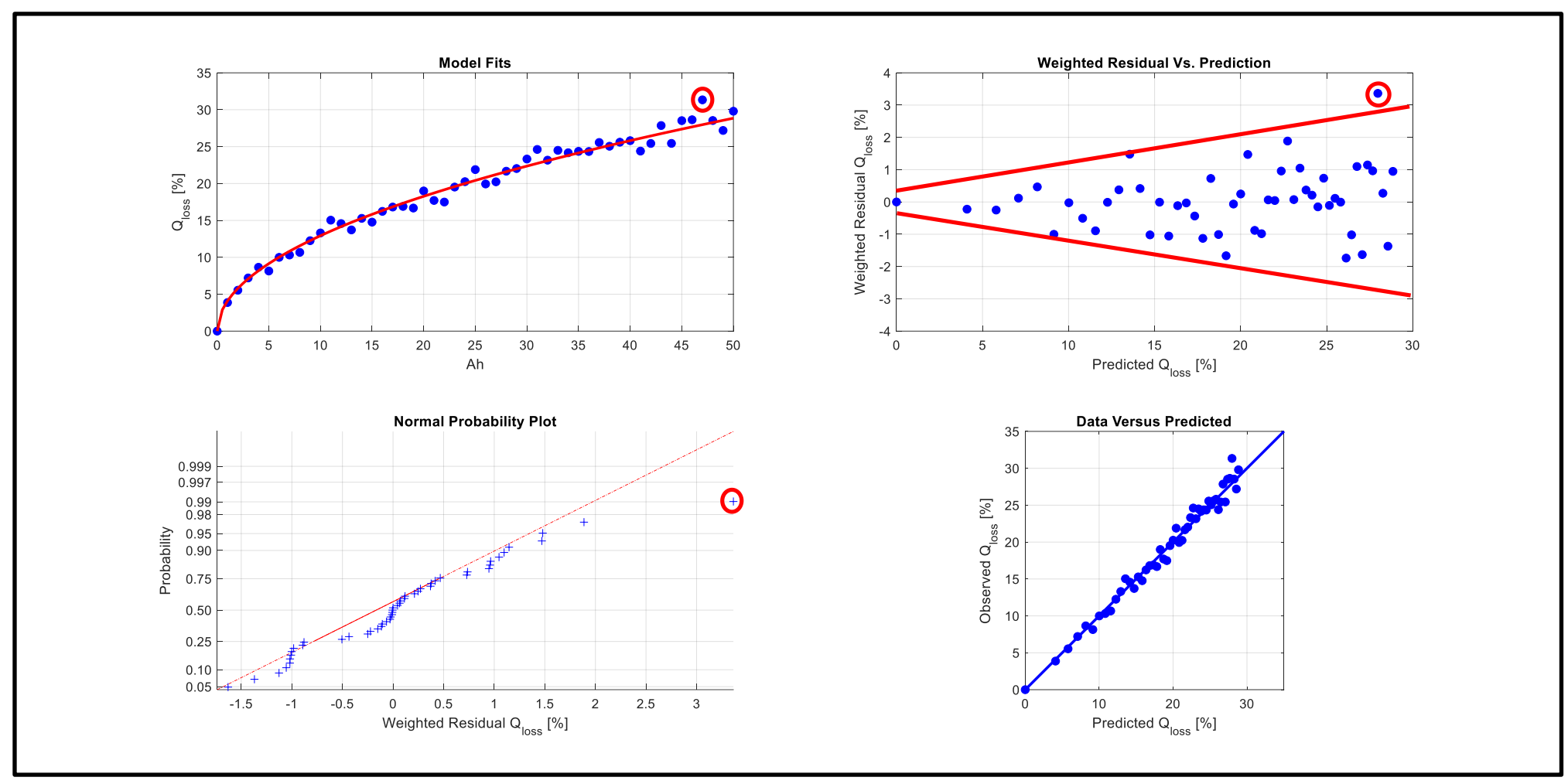

*Figure 1: ROLS analysis for the SoH data. Hyper-parameter optimised using $AIC_c$ formula. The residual versus prediction diagnostic plot (top right) illustrates strong evidence for the presence of heterogeneous variance. There is one outlying observation circled in red. Note, both the independent and dependent data have been mapped to the interval [0, 1]. The optimal value of the hyper-parameter is 0.0074.*

Figure 2 presents diagnostic plots for the equivalent $RIGLS$ fit to the same data. The applied covariance model is of the form $\sigma^2 f(x, \theta)^{2\delta}$. The improvement in the appearance of the weighted residual versus predicted plot is immediately apparent. The estimated value for $\delta$ is 0.973, which is reassuringly close to the *true* value of unity. Further, the optimal value of the hyper-parameter is 0.0190. This is over 3 times larger than the $ROLS$ value. The increased magnitude implies greater bias and yet there is no obvious evidence of this in the diagnostic plots.

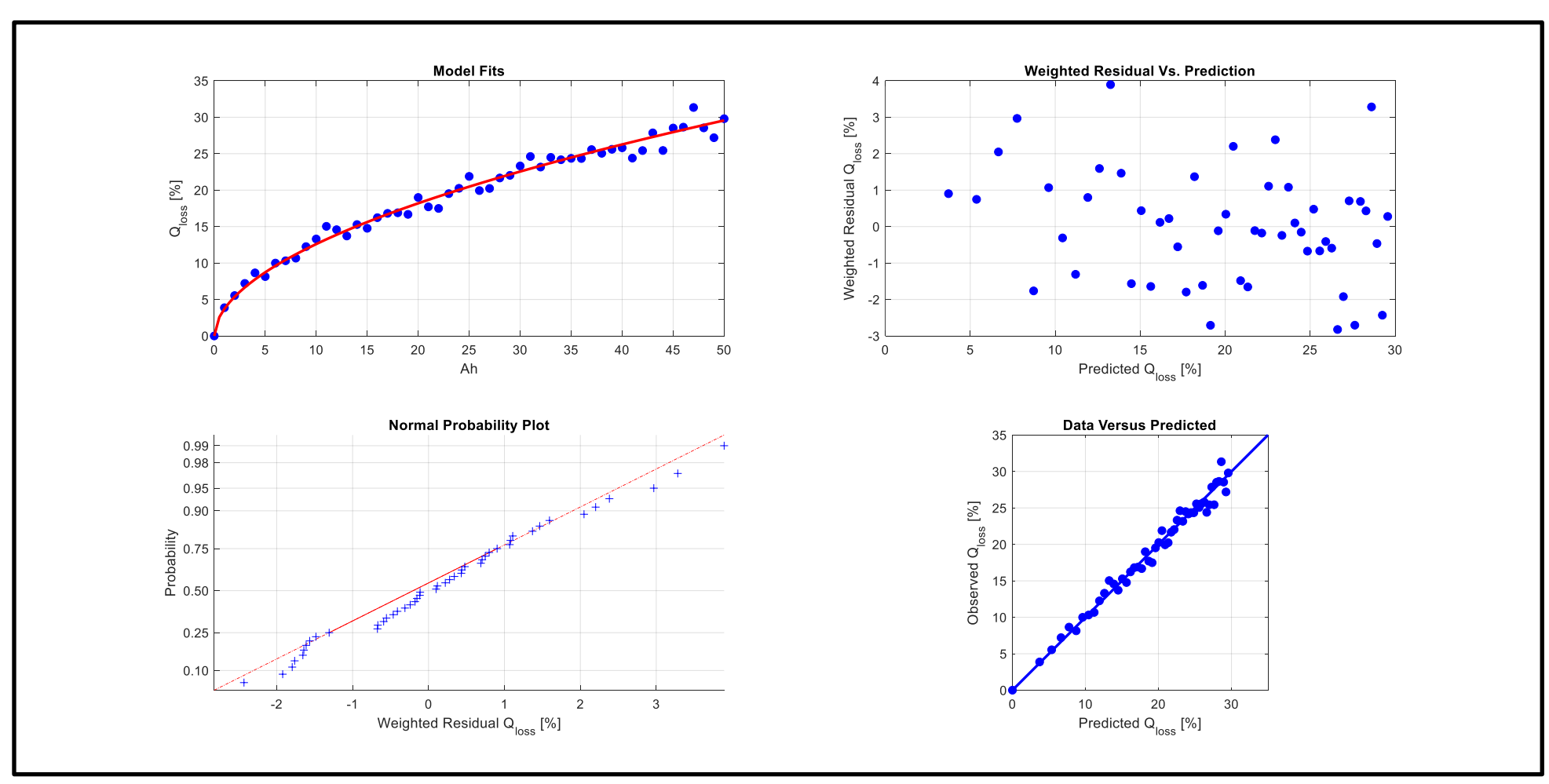

*Figure 2: RIGLS fit to the data. Hyper-parameter optimised using $AIC_c$ formula. The appearance of the weighted residual versus prediction plot is much improved via application of the appropriate covariance model. Both the independent and dependent data have been mapped to the interval [0, 1]. The optimal value of the hyper-parameter is 0.019.*

Finally, Figure 3 presents the diagnostic plots for the $RIGLS$ algorithm, but in this instance optimal hyper-parameter selection is based on $BIC$. These are almost identical in appearance to Figure 2 for the $AIC_c$ optimal $\lambda$-selection case. Here at convergence $\lambda = 0.0456$.

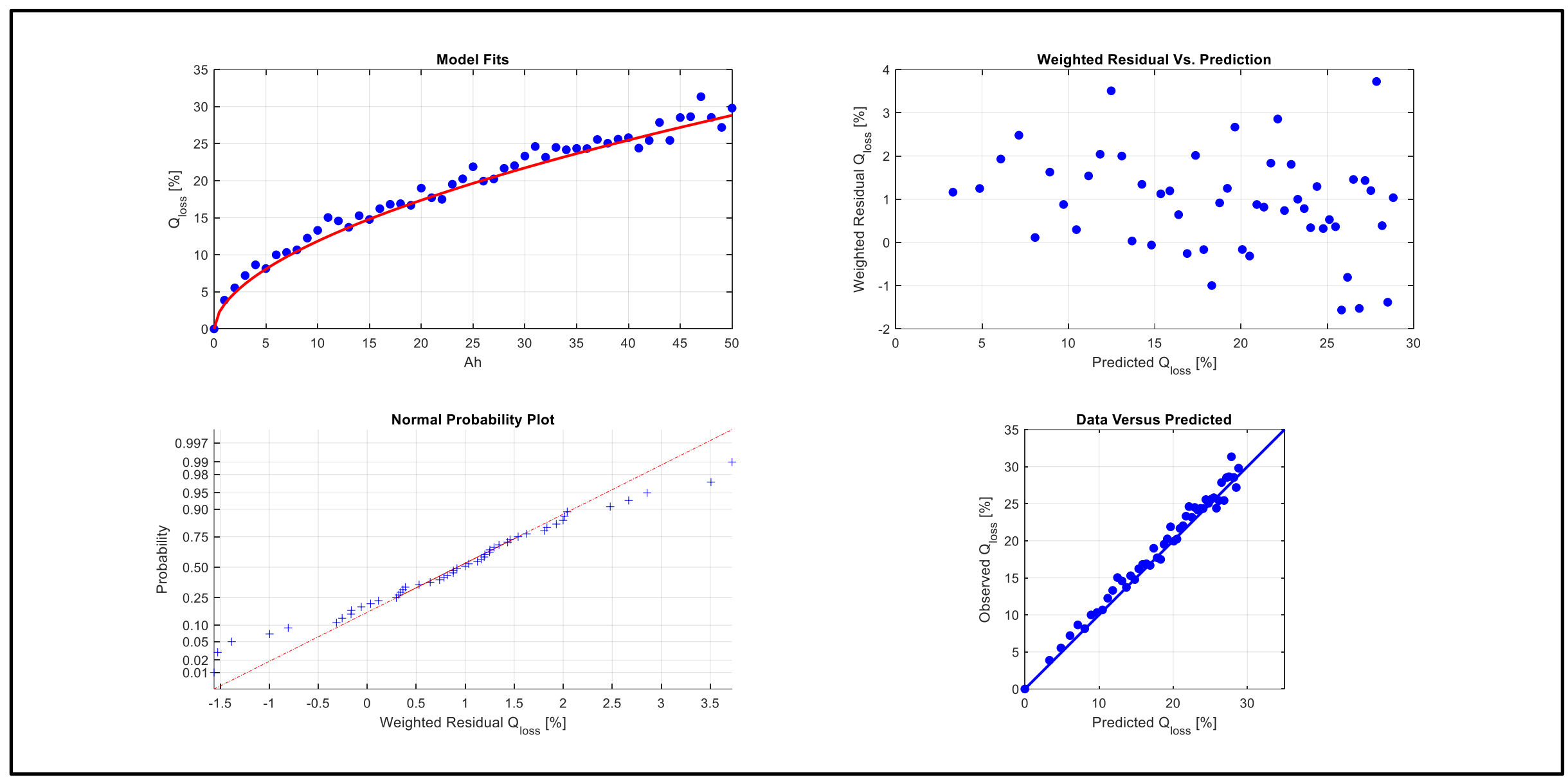

*Figure 3: RIGLS fit to the data. Hyper-parameter optimised using the $BIC$ formula. The appearance of the weighted residual versus prediction plot is much improved via application of the appropriate covariance model Both the independent and dependent data have been mapped to the interval [0, 1]. The optimal value of the hyper-parameter is 0.0456.*

To gain some insight into the impact of the hyper-parameter on the estimates, we calculate the *condition numbers* [xxviii], $\mu_0$ and $\mu_1$, for the matrices $F_0 = (J^T J)$ and $F_1 = (J^T J + \lambda I)$ respectively for all 3 cases. These are presented in Table 1. Recall, the inverse of $F_0$ and $F_1$ are the respective covariance matrices for the $ROLS$ or $RIGLS$ estimates as appropriate.

*Table 1: Condition numbers, optimal hyper-parameter value and degrees of freedom for the various estimates. The ideal condition number is unity. Condition numbers approaching infinity indicate rank degeneracy among the columns/rows of the corresponding matrix.*

| | ROLS | *RIGLS AICc* | *RIGLS BIC* |
|---|---|---|---|
| $\lambda$ | 0.0060 | 0.0190 | 0.0456 |
| $\mu_o$ | 3.0853E+19 | 3.9489E+18 | 2.9766E+18 |
| $\mu_1$ | 2622.6000 | 119.3400 | 630.7300 |
| $\gamma$ | 1.9854 | 1.9906 | 1.9931 |

Regardless of the source, the values of $\mu_o$ are extremely large, indicating that $F_0$ is nearly rank degenerate in all cases. This demonstrates the very poor identifiability of the model, and the ill-posed nature of the unregularized fit. In contrast, despite the relatively small magnitude of the corresponding optimal hyper-parameter values, the associated $\mu_1$ numbers are vastly reduced, with the $AIC_c$ based value exhibiting some superiority. Table 1 also presents the corresponding model degrees of freedom, $\gamma$. In all cases, despite the relatively small associated hyper-parameter values, the degree of freedom reduction is substantial; from 4 to just under 2.

Table 2 presents parameter estimates and 95% confidence intervals for $RIGLS$ estimates. The parameter estimates for the $AIC_c$ and $BIC$ are reasonably close to the *true* values, indicating the induced bias from the hyper-parameter is not large. However, it does though have some practical significance. For example, for the $BIC$ case, the $\beta_1$ estimate has the wrong sign. In a parametric regression study of this nature, where quasi-physical thinking is used to develop the model formulation, incorrect estimation of the sign of the coefficient may lead to inappropriate conclusions regarding the nature and interpretation of the underlying true ageing mechanisms.

Also shown in Table 2 are the 95% confidence intervals for the parameter estimates. Given the inherent poor identifiability of the model structure, the confidence intervals are relatively small, particularly for $\beta_1$. Notice the confidence intervals for $\omega$ and $z$ encompass the true estimate, whereas the intervals for $\beta_1$ and $\beta_2$ do not. This may be symptomatic of the inherent bias induced in regularised estimates, despite the relatively small values of the hyper-parameter. Also, it is interesting to note, despite the larger hyper-parameter value associated with the $BIC$ $RIGLS$ estimator, the corresponding confidence intervals are larger. This follows from the larger condition number for the regularised $F_1$ matrix, resulting in variance inflation of the estimates, demonstrating that the confidence intervals depend on the parameter estimates as well as the size of the hyper-parameter.

*Table 2: Comparison among estimators, parameter values and 95% confidence intervals for the RIGLS estimates.*

| | | RIGLS *AIC$_c$* | | | RIGLS *BIC* | | |
|---|---|---|---|---|---|---|---|
| | **True Value** | LCI | Estimate | UCI | LCI | Estimate | UCI |
| $\omega$ | **0.40** | 0.392 | 0.403 | 0.415 | 0.308 | 0.410 | 0.513 |
| $\beta_1$ | **0.09** | 0.017 | 0.017 | 0.017 | -0.088 | -0.088 | -0.088 |
| $\beta_2$ | **0.02** | 0.054 | 0.056 | 0.058 | 0.088 | 0.105 | 0.123 |
| $\vartheta$ | **0.50** | 0.491 | 0.507 | 0.523 | 0.380 | 0.536 | 0.692 |

# 8 Convergence Analysis of the Optimal Hyper-Parameter Selection Algorithm

Figure 4 presents the output from the first iteration of the algorithm of section 5 versus starting value on the left hand axis, and $|dh(\lambda_0)/d\lambda|$ at the same starting value on the righthand axis. For $\lambda \leq 0.001$ it appears the algorithm can return large negative numbers and appear wildly unstable in places. This is supported by the theorem of section 5, which states that convergence to a unique fixed point requires $|h'(\lambda)| < 1$ in an interval. From the figure, $|h'(\lambda)| \gg 1$ for all $\lambda_0 < 0.001$. For $\lambda \in [0.01,10]$, $|h'(\lambda)| \ll 1$ and rapid convergence of the resulting sequence would be expected. In terms of the number of iterations required, using (22), with $k = 0.20$ and $\lambda_0 = 0.15$, for the interval $\lambda \in [0.1,1]$, for an error bound of 0.00001:

$$r = \frac{log(0.00001) - log(max\{0.05, 0.85\})}{log(0.2)} = 7.0524$$

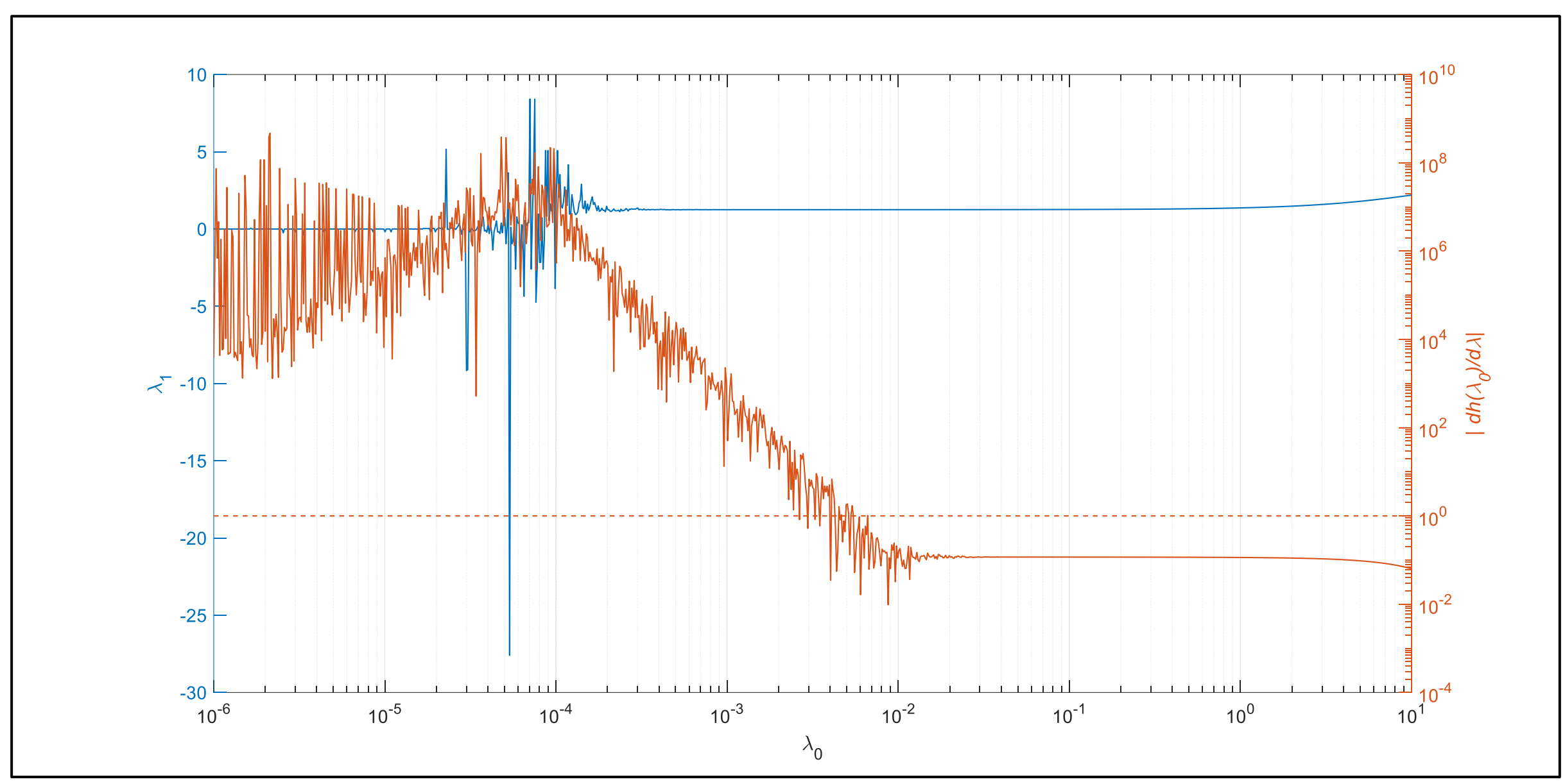

*Figure 4: Convergence Analysis. This plot demonstrates that convergence of the optimal selection algorithm depends on starting position. The dotted line denotes the value of* $|dh(\lambda_0)/d\lambda|$ *below which convergence would be expected.*

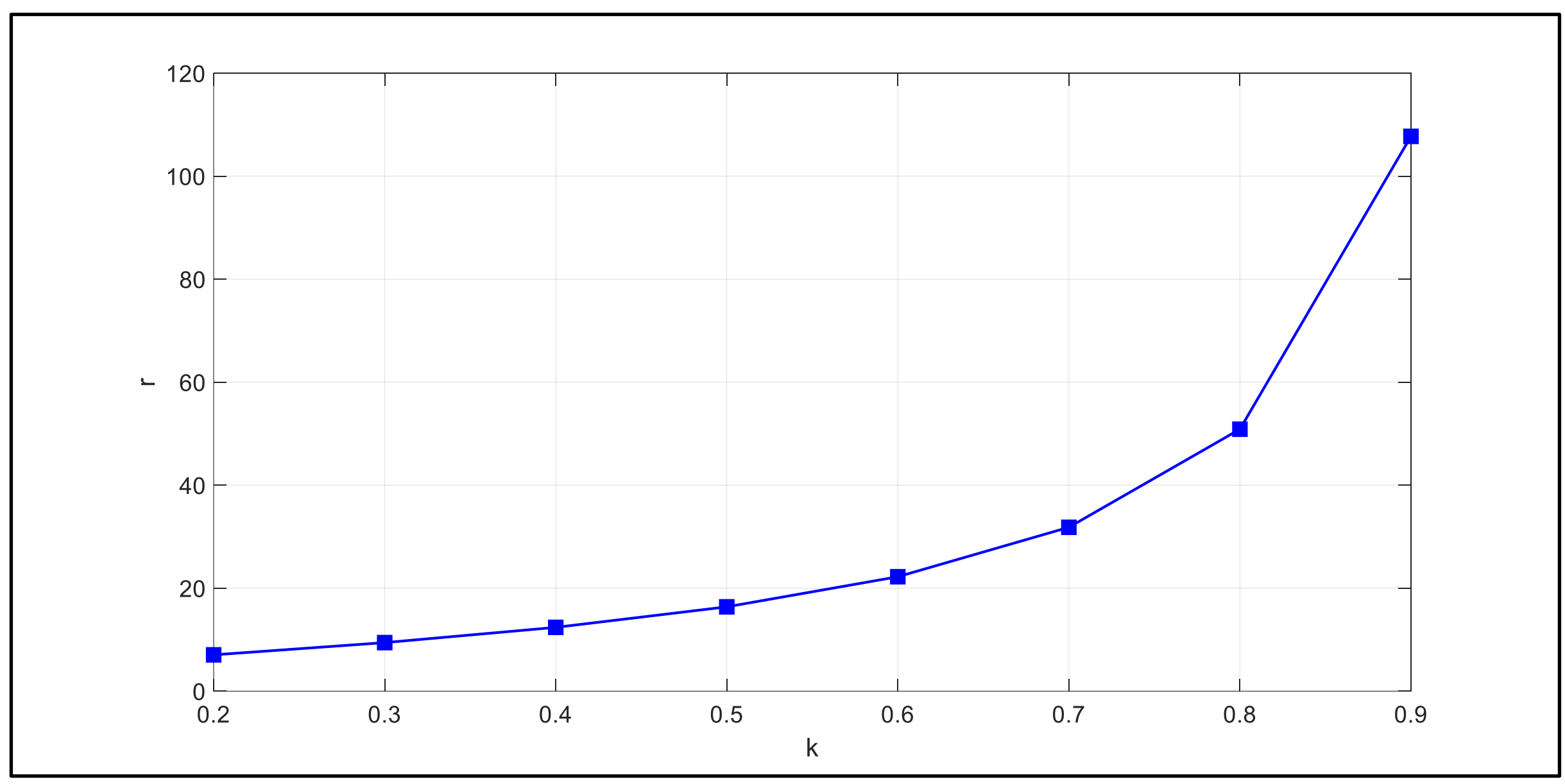

*Figure 5: Expected number of iterations for convergence of the hyper-parameter value with increasing k. In all cases,* $\lambda \in [0.1, 1.0]$ *and* $\lambda_o = 0.15$ *and* $|\lambda_r - \lambda| = 0.00001$.

Hence around 7-8 iterations are required for convergence to the desired error bound. Consider Figure 5, which presents the expected number of iterations of the optimal hyper-parameter algorithm for $k \in [0.2,0.9]$, again with $\lambda_0 = 0.15$ and $\lambda \in [0.1,1]$, again for an error bound of 0.00001. As expected, the number of iterations required for convergence grows exponentially with increasing $k$.

# 9 Conclusions

The optimal hyper-parameter estimation method presented, founded on information theoretic measures, is efficient and converges rapidly given a suitable starting position. Good starting values are any for which $|dh(\lambda_0)/d\lambda|$ is less than unity. However, the smaller the initial $|dh(\lambda_0)/d\lambda|$, the more rapid the convergence. The fixed-point iterative formulae presented are suitable for general covariance structures, including heteroscedastic and or serially correlated errors. When implemented as part of a general $RIGLS$ algorithm scheme, simulation suggests it is possible to preserve quasi-physically motivated structures and return estimates with low bias.

For the simulation study presented, the algorithm performs adequately. $RIGLS$ employing $AIC_c$ as a basis for optimal hyper-parameter estimation marginally outperforms its $BIC$ counterpart in the simulation study presented. Confidence intervals for parameters are relatively narrow, demonstrating the improved precision of the estimates. This follows from the vastly superior condition numbers for the relevant information matrices. However, the simulation study demonstrates that not all confidence intervals reported include the *true* values – an indication of the bias inherent in the regularised approach.

# Appendix A – Derivatives of the Smoother Matrix and its Trace

Here, we provide detailed derivations of the key results, equations (16) through (18). All these equations utilise standard results on matrix derivatives [xxii]. To begin with, the derivative of $S$ with respect to $\lambda$ is straightforward:

We begin by noting:

$$A = Z^T Z + \lambda I$$
$$\Rightarrow Z^T Z = A - \lambda I \qquad \text{(A1)}$$
$$\Rightarrow \frac{\partial A}{\partial \lambda} = I$$

Hence:

$$\frac{\partial S}{\partial \lambda} = Z \frac{\partial A^{-1}}{\partial \lambda} Z^T = -ZA^{-1} \frac{\partial A}{\partial \lambda} A^{-1} Z^T = -ZA^{-2} Z^T \qquad \text{(A2)}$$

Similarly:

$$\frac{\partial S^2}{\partial \lambda} = \frac{\partial}{\partial \lambda}[ZA^{-1}Z^TZA^{-1}Z^T] = \frac{\partial}{\partial \lambda}[ZA^{-1}(A - \lambda I\ )A^{-1}Z^T] = -2ZA^{-2}Z^T + 2\lambda ZA^{-3}Z^T \quad \text{(A3)}$$

Finally, $\frac{\partial \gamma}{\partial \lambda}$ is derived using:

$$\frac{\partial \gamma}{\partial \lambda} = \frac{\partial}{\partial \lambda} tr(ZA^{-1}Z^T) = \frac{\partial}{\partial \lambda} tr(A^{-1}Z^TZ) = \frac{\partial}{\partial \lambda} tr\left(A^{-1}(A - \lambda I)\right) = tr(\lambda A(\lambda)^{-2} - A(\lambda)^{-1}) \quad \text{(A4)}$$

# Appendix B – Fixed Point Iteration Formulae for Hyper-Parameter Estimation

We begin by writing the $BIC$ definition in terms of the profile likelihood (13).

$$BIC = 2L_p + \gamma log(N) \quad \text{(B1)}$$

Differentiating this with respect to $\lambda$ yields:

$$\frac{\partial BIC}{\partial \lambda} = 2\frac{\partial L_p}{\partial \lambda} + log(N)\frac{\partial \gamma}{\partial \lambda} \quad \text{(B2)}$$

Substituting (18) and (19) into (B2):

$$\frac{\partial BIC}{\partial \lambda} = 2\lambda\left(\frac{N}{\sigma(\lambda)^2}\right)\boldsymbol{q}^TZA(\lambda)^{-3}Z^T\boldsymbol{q} + log(N)tr(\lambda A(\lambda)^{-2} - A(\lambda)^{-1}) \quad \text{(B3)}$$

Setting (B3) to 0 and solving for $\lambda$ gives the desired result:

$$\lambda_{BIC} = \frac{\sigma^2 log(N) tr\left(A^{-1} - \lambda A^{-2}\right)}{2N\boldsymbol{q}^TZA^{-3}Z^T\boldsymbol{q}} \quad \text{(B4)}$$

Similarly, writing $AIC_c$ in terms of the profile likelihood:

$$AIC_c = 2L_p + \frac{2\gamma(\gamma+1)}{N-\gamma-1} \quad \text{(B5)}$$

Differentiating (B5) with respect to $\lambda$ and yields:

$$\frac{\partial AIC_c}{\partial \lambda} = 2\frac{\partial L_p}{\partial \lambda} + 2\left[\frac{(2\gamma+1)}{(N-\gamma-1)} - \frac{\gamma(\gamma+1)}{(N-\gamma-1)^2}\right]\frac{\partial \gamma}{\partial \lambda} \tag{B6}$$

Again, setting (B6) to 0 and solving for $\lambda$ yields the desired result:

$$\lambda_{AIC_c} = \frac{\sigma^2\left[\frac{(2\gamma+1)}{(N-\gamma-1)} - \frac{\gamma(\gamma+1)}{(N-\gamma-1)^2}\right]tr\left(A^{-1} - \lambda A^{-2}\right)}{N\boldsymbol{q}^T Z A^{-3} Z^T \boldsymbol{q}} \tag{B7}$$

## Appendix C – Derivative of the Regularised Cost Function

The appropriate regularised cost function for ridge regression is:

$$R = \frac{1}{2}\big(y - f(\theta)\big)^T W^{-1}(\theta_r,\ \delta_{r+1}\ )\big(y - f(\theta)\big) + \frac{\lambda}{2}\theta^T\theta \tag{C1}$$

Expanding the first term on the righthand side of (C1):

$$R = \frac{1}{2}\big(y^T W^{-1} y - y^T W^{-1} f(\theta) - f(\theta)^T W^{-1} y + f(\theta)^T W^{-1} f(\theta)\big) + \frac{\lambda}{2}\theta^T\theta \tag{C2}$$

Differentiating (C2) with respect to $\theta$ and collecting terms yields the desired result:

$$\frac{\partial R}{\partial \theta} = J^T(\theta) W^{-1}\big(f(x,\theta) - y\big) + \lambda\theta \tag{C3}$$

[xxiii] Cary, M., A Model Based Methodology for the Calibration of a Port Fuel Injection, Spark-Ignition Engine, *PhD Thesis*, University of Bradford, 2003.

[xxiv] Burden, R. L., Faires, J. D., Reynolds, A. C., Numerical Analysis, *Wadsworth International Student Edition*, *Second Edition*, 1981.

[xxv] Abramowitz M, Stegun I, (1972), Handbook of Mathematical Functions with Formulas, Graphs, and Mathematical Table, *Dover Publications*

[xxvi] Davidian, M., Giltinan, D. M., Nonlinear Models for Repeated Measurement Data, *Chapman & Hall*, First Edition, 1995.

[xxvii] Gamma, E., Helm, R., Johnson, R., Vlissides, J., Design Patterns, Elements of Reusable Object-Oriented Software, *Addison – Wesley*, 1995.

[xxviii] Gill, P. E., Murray, W., Wright, M. H., Practical Optimization, Academic Press, First Edition, 1981.